\documentclass[reprint,superscriptaddress,amsmath,amssymb,aps,prb]{revtex4-2}
\usepackage{color}
\usepackage{graphicx}
\usepackage{dcolumn}
\usepackage{bm}
\usepackage{hyperref}
\usepackage{tabularx}
\usepackage[T1]{fontenc}
\usepackage{pifont}
\newcommand{\cmark}{\ding{51}}%
\newcommand{\xmark}{\ding{55}}%
\usepackage{array}
\newcolumntype{P}[1]{>{\centering\arraybackslash}p{#1}}
\makeatletter

\begin{document}

\title{Bridging Theory and Experiment in Materials Discovery: Machine-Learning-Assisted Prediction of Synthesizable Structures}

\author{Yu Xin}
\altaffiliation{These authors contributed equally to this work}
\affiliation{Key Laboratory of Material Simulation Methods and Software of Ministry of Education, College of Physics, Jilin University, Changchun 130012, China}

\author{Peng Liu}
\altaffiliation{These authors contributed equally to this work}
\affiliation{Key Laboratory of Material Simulation Methods and Software of Ministry of Education, College of Physics, Jilin University, Changchun 130012, China}
\affiliation{State Key Laboratory of High Pressure and Superhard Materials, College of Physics, Jilin University, Changchun 130012, China}
\affiliation{Laboratory of Computational Materials Physics, College of Physics and Communication Electronics, Jiangxi Normal University, Nanchang 330022, China}

\author{Zhuohang Xie}
\affiliation{Key Laboratory of Material Simulation Methods and Software of Ministry of Education, College of Physics, Jilin University, Changchun 130012, China}
\affiliation{International Center of Future Science, Jilin University, Changchun 130012, China}

\author{Wenhui Mi}
\affiliation{Key Laboratory of Material Simulation Methods and Software of Ministry of Education, College of Physics, Jilin University, Changchun 130012, China}

\author{Pengyue Gao}
\affiliation{Key Laboratory of Material Simulation Methods and Software of Ministry of Education, College of Physics, Jilin University, Changchun 130012, China}

\author{Hong Jian Zhao}
\email{physzhaohj@jlu.edu.cn}
\affiliation{Key Laboratory of Material Simulation Methods and Software of Ministry of Education, College of Physics, Jilin University, Changchun 130012, China}

\author{Jian Lv}
\email{lvjian@jlu.edu.cn}
\affiliation{Key Laboratory of Material Simulation Methods and Software of Ministry of Education, College of Physics, Jilin University, Changchun 130012, China}

\author{Yanchao Wang}
\email{wyc@calypso.cn}
\affiliation{Key Laboratory of Material Simulation Methods and Software of Ministry of Education, College of Physics, Jilin University, Changchun 130012, China}
\affiliation{State Key Laboratory of High Pressure and Superhard Materials
, College of Physics, Jilin University, Changchun 130012, China}

\author{Yanming Ma}
\email{mym@jlu.edu.cn}
\affiliation{Key Laboratory of Material Simulation Methods and Software of Ministry of Education, College of Physics, Jilin University, Changchun 130012, China}
\affiliation{State Key Laboratory of High Pressure and Superhard Materials
, College of Physics, Jilin University, Changchun 130012, China}
\affiliation{International Center of Future Science, Jilin University, Changchun 130012, China}
\date{\today}

\begin{abstract}
Even though thermodynamic energy-based crystal structure prediction (CSP) has revolutionized materials discovery, the energy-driven CSP approaches often struggle to identify experimentally realizable metastable materials synthesized through kinetically controlled pathways, creating a critical gap between theoretical predictions and experimental synthesis.  
Here, we propose a synthesizability-driven CSP framework that integrates symmetry-guided structure derivation with a Wyckoff encode-based machine-learning model, allowing for the efficient localization of subspaces likely to yield highly synthesizable structures. 
Within the identified promising subspaces, a structure-based synthesizability evaluation model, fine-tuned using recently synthesized structures to enhance predictive accuracy, is employed in conjunction with \textit{ab initio} calculations to systematically identify synthesizable candidates.
The framework successfully reproduces 13 experimentally known XSe (X = Sc, Ti, Mn, Fe, Ni, Cu, Zn) structures, demonstrating its effectiveness in predicting synthesizable structures.
Notably, 92,310 structures are filtered from the 554,054 candidates predicted by GNoME, exhibiting great potential for promising synthesizability. Additionally, eight thermodynamically favorable Hf-X-O (X = Ti, V, and Mn) structures have been identified, among which three HfV$_2$O$_7$ candidates exhibit high synthesizability, presenting viable candidates for experimental realization and potentially associated with experimentally observed temperature-induced phase transitions.
This work establishes a data-driven paradigm for machine-learning-assisted inorganic materials synthesis, highlighting its potential to bridge the gap between computational predictions and experimental realization while unlocking new opportunities for the targeted discovery of novel functional materials.
\end{abstract}

\maketitle
\section{Introduction}\label{sec_intro}
Advancements in crystal structure prediction (CSP), where thermodynamic energy-based stability predictions have traditionally been used to estimate synthesizability, offer significant opportunities to accelerate the discovery of new materials~\cite{woodley2008crystal,oganov2011evolutionary, oganov2018crystal,wang2022crystal}.
This is exemplified by the theory-driven discovery of high-temperature superconducting superhydrides, including CaH$_6$, YH$_9$, and LaH$_{10}$, which exhibit critical temperatures of 215 K, 243 K, and 260 K, respectively~\cite{sun2024clathrate}. 
However, a major bottleneck in these approaches is that many computationally designed materials, despite being thermodynamically stable, are not synthesizable~\cite{lee2022machine}. As a result, only a small fraction of the candidates predicted by CSP are successfully synthesized.

\begin{figure*}[ht]
    \centering
    \includegraphics[width=1\linewidth]{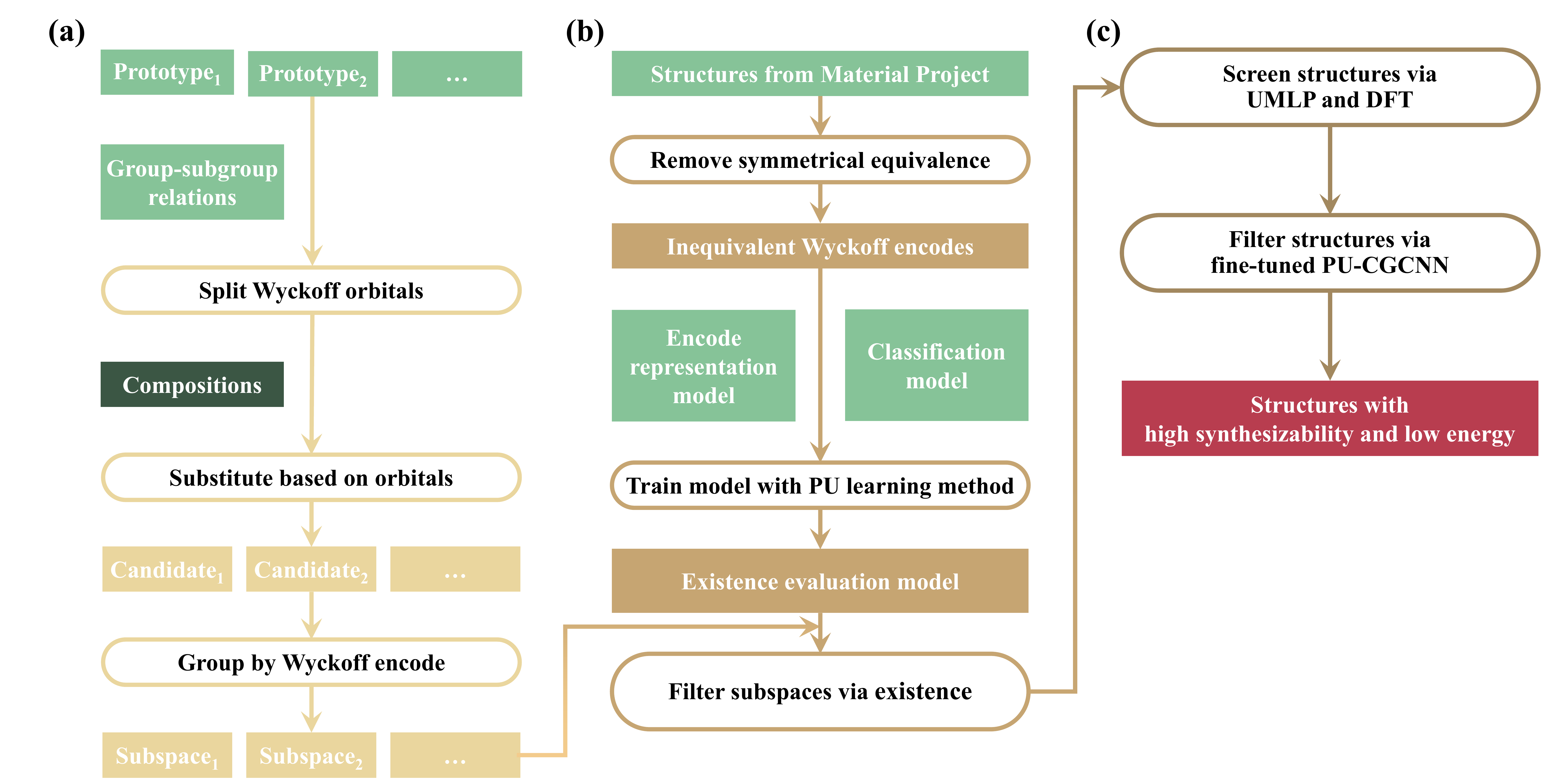}
    \caption{
    The workflow of the synthesizability-driven CSP method.
    The left-middle panel illustrates the structure derivation process, the middle panel outlines the training of the Wyckoff encode-based existence evaluation model, and the right panel showcases the candidate screening process using universal machine learning potential (UMLP) \cite{dunn2020benchmarking,focassio2024performance}, density functional theory (DFT) \cite{martin2020electronic}, and the structure-based synthesizability evaluation model, fine-tuned on the  Positive-Unlabeled Crystal Graph Convolutional Neural Networks (PU-CGCNN)~\cite{jang2020structure}.}
    \label{fig: workflow}
\end{figure*}

To accelerate the experimental discovery of new materials, the synthesizability of candidate structures must be considered to bridge the gap between theoretical predictions and experimental synthesis~\cite{moosavi2020role}. This marks the emergence of a new paradigm in which computational research not only predicts materials with unexpected properties but also offers valuable insights to guide experimental synthesis~\cite{jang2020structure,wang2022ulsa,mcdermott2021graph,aykol2021rational,chen2023matchat,he2023precursor}. Nevertheless, accurately predicting the synthesizability of structures in computational models remains a formidable challenge due to the intrinsic complexity of the synthesis process. This complexity arises from various factors, including synthesis routes, thermodynamic conditions, and other experimental parameters—most of which cannot be fully predicted based on thermodynamic or kinetic constraints alone~\cite{davariashtiyani2021predicting}.

In the field of inorganic materials, the growing availability of crystal structure databases~\cite{jain2013commentary, belsky2002new, zagorac2019recent} and advancements in machine learning algorithms offer a promising pathway to addressing the synthesizability challenge~\cite{ramprasad2017machine, liu2017materials, schmidt2019recent, damewood2023representations, ryan2018crystal, aykol2019network}. Several machine-learning-based approaches have been developed to predict the synthesizability of materials~\cite{lee2022machine, antoniuk2023predicting, jang2024synthesizability, kim2024large, jang2020structure, gu2022perovskite, gleaves2023materials, amariamir2024syncotrain, sun2025crystal, davariashtiyani2021predicting, zhu2023predicting, song2024large, kim2024explainable}. These approaches, based solely on composition or crystal structure, can be classified into two categories. The first category employs composition embeddings to construct a classification model for predicting the synthesizability of materials~\cite{lee2022machine, antoniuk2023predicting, jang2024synthesizability, kim2024large}. For example, Antoniuk \textit{et al}. encode composition using a 94-dimensional vector, which serves as input for a classification model to estimate the likelihood of synthesizing inorganic materials~\cite{cubuk2019screening, zhou2018learning}. However, composition alone is insufficient for CSP, as demonstrated by universally occurring polymorphs such as diamond and graphite.

The second category integrates various structural representations with semi-supervised machine learning techniques, such as positive-unlabeled learning, to predict whether a structure with a specific atomic arrangement can be synthesized, without relying on thermodynamic metrics. These structural representations encompass graph-based encoding, three-dimensional pixel-wise images of crystal structures, and Fourier-transformed crystal features that integrate both real-space and reciprocal-space information~\cite{jang2020structure, gu2022perovskite, gleaves2023materials, amariamir2024syncotrain, sun2025crystal, davariashtiyani2021predicting, zhu2023predicting}. Additionally, large language models have been fine-tuned to construct classification models for synthesizability prediction by encoding crystal structures into textualized crystallographic information files that retain only the representative positions of Wyckoff orbitals~\cite{song2024large} or by leveraging Robocrystallographer to generate descriptive, text-based summaries of crystal structures~\cite{ganose2019robocrystallographer, kim2024explainable}.

These models have advanced the field of synthesizability-driven CSP. However, most existing structure-based synthesizability evaluation models are trained on either a limited set of experimental structures or structures near local minima of the potential energy surface, thereby restricting their broader applicability in CSP. In particular, synthesizability is a discrete, binary property that lacks a well-defined derivative concerning structural configuration, rendering many efficient local optimization techniques commonly used in energy-driven CSP inapplicable. These limitations exacerbate the challenges in ensuring the transferability of the synthesizability-driven CSP method.

Here, a synthesizability-driven CSP approach is proposed to guide the synthesis of inorganic structures using only predefined elemental stoichiometry. This approach integrates structure derivation via group-subgroup relations from synthesized prototypes with a fine-tuned, structure-based synthesizability evaluation model. In particular, a symmetry-oriented divide-and-conquer strategy employing Wyckoff encode~\cite{goodall2022rapid} is developed to efficiently identify the most promising regions of the configuration space with a high probability of yielding synthesizable structures, rather than exhaustively searching the entire potential energy surface, thus significantly enhancing the efficiency of CSP in discovering experimentally accessible materials. This approach identifies 92,310 synthesizable structures from the 554,054 candidates predicted by the Graph Networks for Materials Science (GNoME)~\cite{merchant2023scaling} and is successfully applied to reproduce 13 experimentally synthesized XSe (X = Sc, Ti, Mn, Fe, Ni, Cu, Zn) structures. Notably, three novel HfV$_2$O$_7$ phases are predicted, exhibiting low formation energies and high synthesizability. These predictions serve as proofs of concept, demonstrating the substantial potential of data-driven learning in bridging the gap between theoretical modeling and experimental synthesis. This work highlights both the feasibility and the remarkable effectiveness of the synthesizability-driven CSP approach in accelerating the discovery of inorganic crystal materials.

\begin{figure*}[ht]
    \centering
    \includegraphics[width=0.865\linewidth]{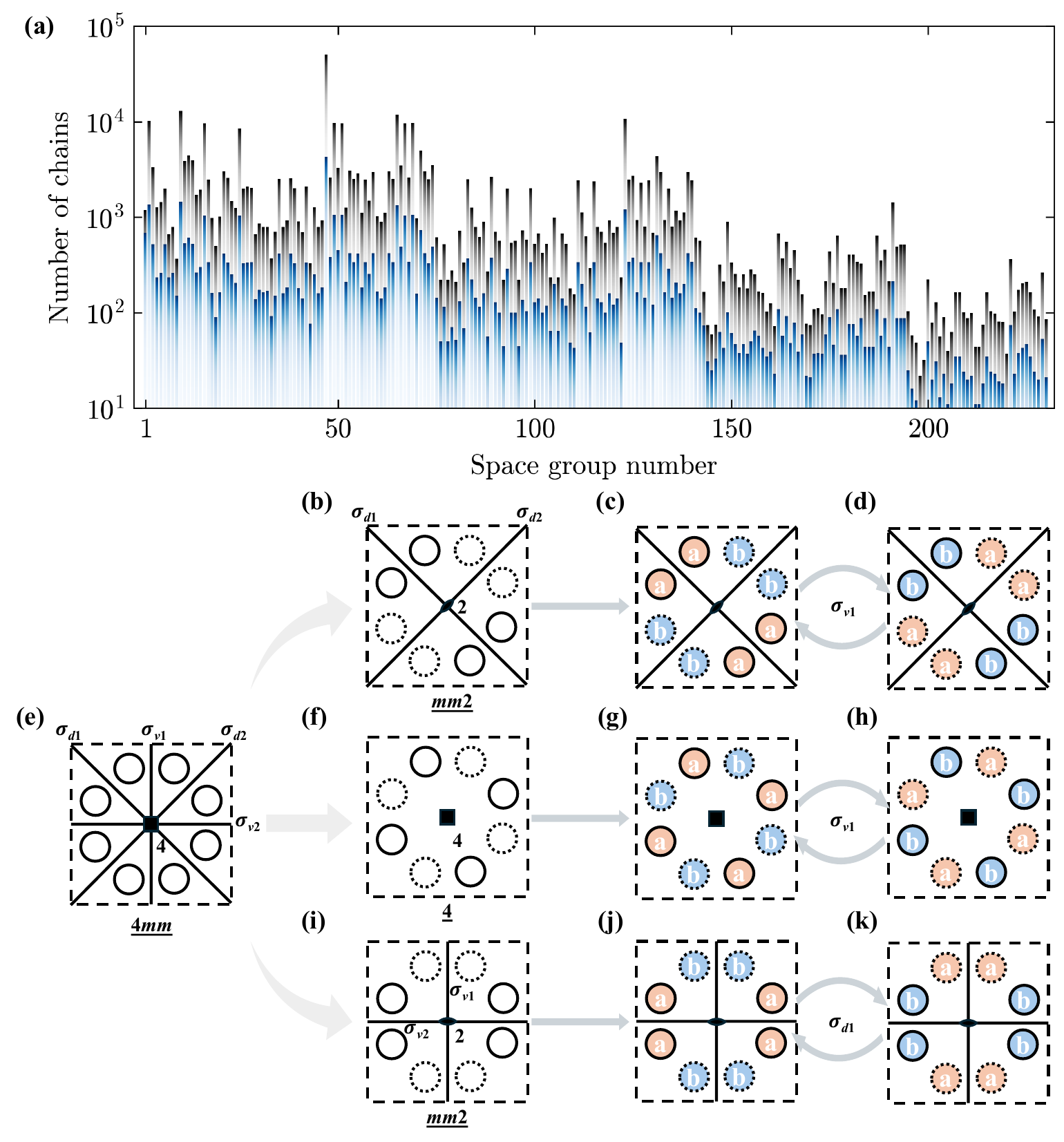}
    \caption{
    (a) The number of all and inequivalent group-subgroup transformation chains with a subgroup index less than or equal to 10 for the space groups. The height of the dark bar represents the number of all the group-subgroup transformation chains, while the blue bar demonstrates the number of inequivalent ones.
    (b-k) The schematic representation of the substitutional structure derivation process.
    (e) depicts a parent structure with \(4mm\) symmetry, characterized by eight symmetry operations: \(1\), \(2\), \(4^{+}\), \(4^{-}\) and four mirror planes oriented along [01], [-10], [-1-1], and [-11], denoted as \(\sigma_{v2}\), \(\sigma_{v1}\), \(\sigma_{d1}\), and \(\sigma_{d2}\), respectively. The circles in (e) represent a Wyckoff orbital with a multiplicity of 8.
    The uncolored diagrams (b, f, i) illustrate three orbital splitting patterns corresponding to three subgroups with index 2. 
    The original orbital splits into two 4-fold orbitals, depicted by circles with solid and dashed boundaries.
    Both (b) and (i) exhibit \(mm2\) symmetry with different symmetric elements, while (f) retains \(4\) symmetry. 
    (c) and (d) illustrate two substitution styles derived from (b), which are equivalent under \(\sigma_{v1}\).  
    (g) and (h) depict two substitution styles derived from (f), which are also equivalent under \(\sigma_{v1}\).  
    (j) and (k) show two substitution styles derived from (i), which are equivalent under \(\sigma_{d1}\).
    }
    \label{fig: workflow detail 2}
\end{figure*}

\begin{figure*}[ht]
    \centering
    \includegraphics[width=1\linewidth]{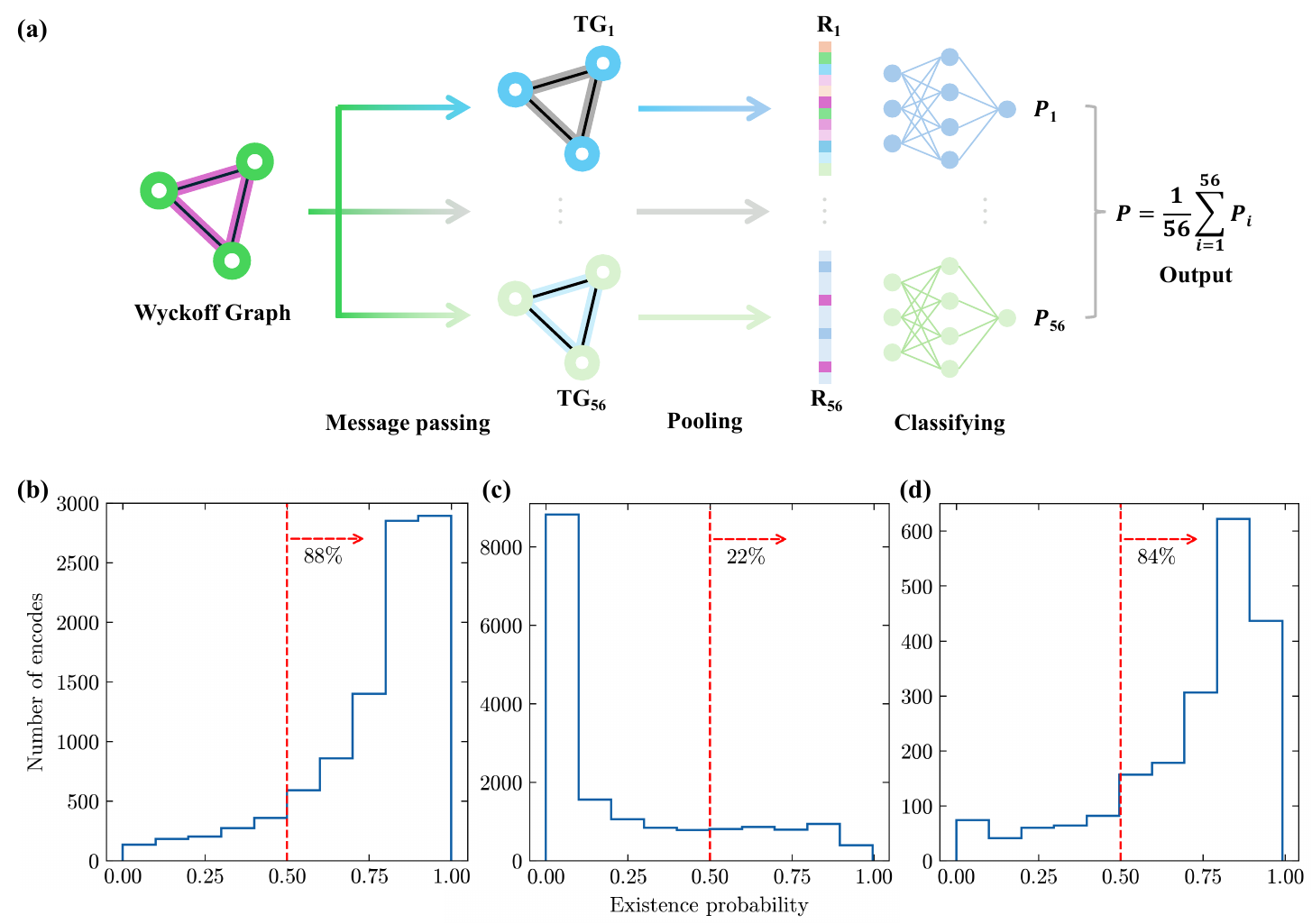}
    \caption{
    (a) The schematic diagram of the Wyckoff encode-based existence probability evaluation model.  
    Each node in the input Wyckoff graph is characterized by the space group number, occupied Wyckoff orbital details, element type, and lattice type.
    Each row represents an independent Wyckoff encode-based existence probability evaluation model.  
    The initial Wyckoff graph is transformed into a new graph (TG\(_i \)) through a message-passing network, followed by pooling to generate a latent representation of (R\(_i \)). A standard classification network is used to assess the existence probability of \( P_i \) from R\(_i \).  
    The final existence probability score ($P$) for each input graph or Wyckoff encode is evaluated by averaging the probabilities (\( P_i \)) predicted by all independent models. (b-d) Performance evaluation of the Wyckoff encode-based existence probability model on encodes derived from synthesized and unsynthesized structures in the test set, as well as recently synthesized structures from the Inorganic Crystal Structure Database (ICSD)~\cite{zagorac2019recent, belsky2002new} that are not included in the training dataset.
}
    \label{fig: workflow detail 1}
\end{figure*}

\section{Method}
The workflow of our synthesizability-driven CSP framework, as illustrated in Fig.~\ref{fig: workflow}, consists of three key steps. First, structure derivation via group-subgroup relations is employed to generate structures from synthesized prototypes, ensuring that the sampled structures retain the atomic spatial arrangement of experimentally realized ones. Second, these structures are classified into distinct configuration subspaces, labeled by Wyckoff encodes. The subspaces are then filtered based on the probability of synthesizable structures existing within each subspace, as predicted by our machine-learning model. Finally, structural relaxations are applied to all structures in the selected subspaces, followed by synthesizability evaluations, ultimately yielding low-energy, high-synthesizability candidates.

\subsection{Structure derivation via group-subgroup relations from synthesized structures\label{sosdm}}

In general, the accuracy and transferability of synthesizability evaluation models depend critically on the training set. However, due to the limited availability of synthesized structures for structure-based synthesizability evaluation models, general and accurate predictive models capable of assessing synthesizability across diverse crystal structure types and chemical compositions remain lacking. In particular, the extensive size and intrinsic uncertainty of the sample space of unsynthesized crystals pose a significant challenge for synthesizability prediction. As a result, the strategy of randomly generating structures with a given composition, typically employed in traditional CSP methods~\cite{wang2010crystal, wangCALYPSOMethodCrystal2012, shao2022symmetry}, is inapplicable to the synthesizability-driven CSP approach, as shown in Fig. S2. 
To enhance both prediction accuracy and confidence in ranking synthesizable structures, we propose a symmetry-guided structure derivation method based on group-subgroup relations, which differs from previous approaches~\cite{ferreira1991stabilityFWZ, grau2007symmetrySOD, hart2008algorithmEnumlib, hart2009generatingEnumlib, hart2012generatingEnumlib, morgan2017generatingEnumlib, mustapha2013useCrystal, d2013symmetryCrystal, okhotnikov2016supercell, lian2022highlyDisorder, lian2020algorithmDisorder, he2021biasedSAGAR, prayogo2022Shry}. This approach systematically derives candidate structures directly from synthesized prototypes, ensuring that the generated structures are highly relevant to experimentally realizable materials within our synthesizability-driven CSP framework (Fig.~\ref{fig: workflow}(a)).

This method consists of three key steps: constructing the prototype database, identifying symmetry-inequivalent group-subgroup transformation chains, and performing element substitution. 
First, prototype structures are derived from synthesized structures in the Materials Project (MP) database~\cite{jain2013commentary}, which are standardized by discarding atomic species to restore the highest possible symmetry in their spatial arrangements. Redundant structures are then removed using the coordination characterization function~\cite{su2017construction}, yielding a final set of 13,426 prototype structures.

Second, group-subgroup transformation chains are constructed to systematically describe the symmetry reduction of the prototype, progressing from maximal subgroups to lower subgroups in increasing index $i$. These transformation chains form the foundation for our structure derivation method. As the maximal subgroups of each space group are well documented in the International Tables for Crystallography~\cite{hahn2005international}, any group-subgroup transformation chain can be systematically determined using a graph-based approach, as implemented in the SUBGROUPGRAPH~\cite{ivantchev2000subgroupgraph}. For further details on the construction of group-subgroup transformation chains, see the Supporting Information  (SI), Sec. I.

In general, derivative structures generated from conjugate subgroups are crystallographically equivalent~\cite{muller2013symmetry}, necessitating the elimination of redundant subgroups. Given two subgroups \(\mathcal{H}_i\) and \(\mathcal{H}_j\) of \(\mathcal{G}\), if there exists an element \(g \in \mathcal{G}\) such that \(g^{-1}\mathcal{H}_j g = \mathcal{H}_i\), then \(\mathcal{H}_i\) and \(\mathcal{H}_j\) are conjugate subgroups. Consequently, eliminating conjugate subgroups prevents redundancy in derivative structure generation. Statistical analysis reveals that, except for \( P1 \), more than 50$\%$ of all possible transformation chains with \( i \leq 10 \) are eliminated by filtering out conjugate subgroups in most space groups. Notably, in the \( Pmmm \) space group, 46,434 out of 50,698 chains (92\%) are removed, as shown in Fig.~\ref{fig: workflow detail 2}(a) and Fig. S1.

Third, the group-subgroup transformation chains are leveraged to guide element substitution based on the target composition, as illustrated in Fig.~\ref{fig: workflow}(a). During this process, the Wyckoff positions of the parent prototype split into distinct patterns, leading to symmetrically inequivalent transformation chains of the same index.

As illustrated in Figs.~\ref{fig: workflow detail 2}(b, f, i), an 8-fold Wyckoff orbital in the parent structure with \(4mm\) symmetry (Fig.~\ref{fig: workflow detail 2}(e)) splits into two distinct orbitals when either mirror symmetry or 4-fold rotational symmetry is removed. 
Following the group-subgroup transformation chain, Wyckoff orbital splitting enables element substitution in an orbital-wise manner to achieve the target composition and target subgroup symmetry. For example, a 1:1 composition yields two possible substitution patterns for a derivative structure with \(4\) symmetry, a principle that also applies to transformation chains leading to \(mm2\) symmetry.

However, two types of duplication in orbital-wise element substitution must be identified and removed: (1) substitutions that produce a structure with higher symmetry than the intended subgroup, and (2) substitution styles that are interchanged by a symmetry operation \( g \in \mathcal{G} \) but \( g \notin \mathcal{H} \), where \(\mathcal{G}\) is the space group of the parent structure and \(\mathcal{H}\) is the subgroup after symmetry reduction, the details of duplication removal are illustrated in the SI Sec. I.

\subsection{Wyckoff encode-based existence probability filtering\label{wesem}}

Due to the vast configurational space of potential materials, an effective learning strategy is urgently required in synthesizability-driven CSP to efficiently identify the subspaces containing synthesizable structures while disregarding specific atomic positions. 
Although composition-based methods~\cite{lee2022machine, antoniuk2023predicting, jang2024synthesizability, kim2024large} attempt to tackle this issue, they remain insufficient for CSP, as they discard all structural information. 
Here, the Wyckoff encode~\cite{goodall2022rapid} is introduced to represent crystal structures while preserving symmetry information. It consists of five segments: the reduced stoichiometry, the Pearson symbol (crystal family, centering type, and the number of atoms in the conventional cell), the space group number, the Wyckoff orbitals occupied by each element, and the element symbols. 
For example, the Wyckoff encode for sodium chloride, AB\_cF8\_225\_a\_b:Na-Cl, indicates that Na atoms occupy the \( 4a \) Wyckoff orbital and Cl atoms occupy the \( 4b \) orbital of space group \( Fm\overline{3}m \) (the 225$^{th}$ space group), with a face-centered cubic unit cell. Each encode defines a subspace within the configuration space, which is spanned by lattice parameters and Wyckoff orbital variables.

\begin{figure*}[ht]
    \centering
    \includegraphics[width=1\linewidth]{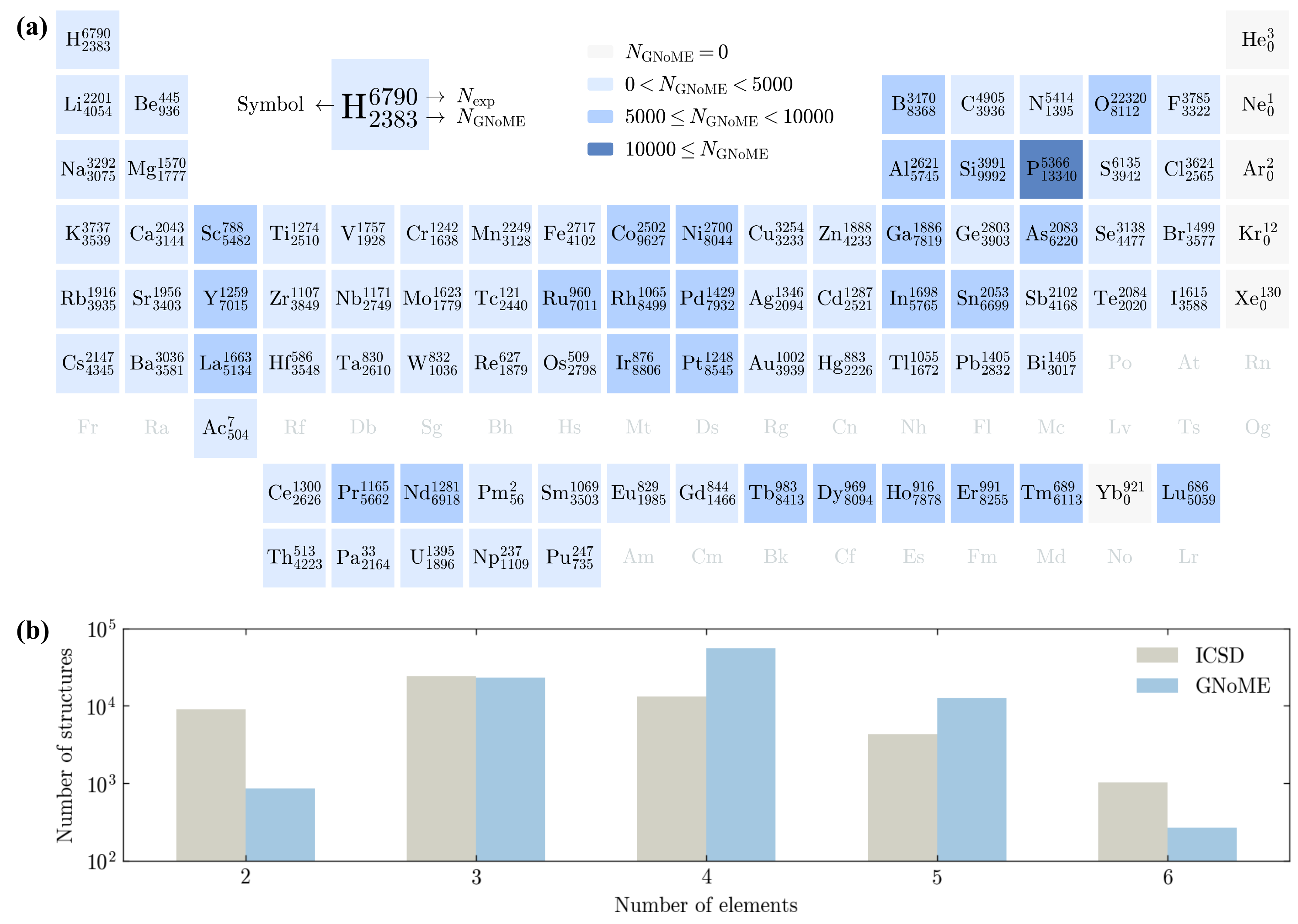}
    \caption{(a) Statistics on the elemental distribution of synthesized and synthesizable structures from GNoME materials exploration. $N_{\mathrm{exp}}$ denotes the number of experimentally synthesized structures containing a given element. $N_{\mathrm{GNoME}}$ represents the number of synthesizable structures—defined as those with $P > 0.5$ and CLscore $> 0.5$—from GNoME exploration that contain the element. Structures containing gray-colored highly radioactive elements are excluded from the statistics. (b) Number of structures as a function of the number of constituent elements, considering synthesized structures and synthesizable structures from the GNoME materials exploration.}
    \label{fig: gnome_icsd}
\end{figure*}

\begin{table*}[ht]
    \centering
    \caption{GNoME-identified candidates that match the experimentally synthesized structures in the ICSD are presented. $P$ denotes the existence probability predicted by the Wyckoff encode-based model, while CLscore reflects the structural synthesizability evaluated by the fine-tuned PU-CGCNN model. Structures that satisfy both $P > 0.5$ and CLscore $> 0.5$ are labeled as success (\cmark), whereas others are labeled as unsuccess (\xmark).}
    \label{tab: gnome_icsd}
    \begin{tabularx}{1\linewidth}{P{2.2cm}P{7.3cm}P{1.2cm}P{1.5cm}P{2cm}P{1.8cm}P{1cm}}
\hline\hline  
GNoME ID & Wyckoff encode & $P$ & CLscore & ICSD ID & Reproduction & Ref.\\
771ca50678&A3B3C\_oP28\_62\_3c\_3c\_c:As-Ca-In&0.92&0.88&icsd-126519&\cmark&\cite{peng2022synthesis}\\
29cb76f211&AB2C6D\_mP20\_4\_a\_2a\_6a\_a:K-P-S-Tb&0.90&0.96&icsd-175473&\cmark&\cite{zhao2024krep2s6}\\
ac0d056eef&ABC2D6\_mP20\_4\_a\_a\_2a\_6a:Dy-K-P-S&0.90&0.96&icsd-175470&\cmark&\cite{zhao2024krep2s6}\\
8b8d73e608&A2BC2D6\_mP22\_14\_e\_a\_e\_3e:K-Mg-P-S&0.89&0.73&icsd-170407&\cmark&\cite{zhao2023noncentrosymmetric}\\
c3b3bce217&A2B2C2D7\_mP52\_14\_2e\_2e\_2e\_7e:B-K-Nd-O&0.86&0.80&icsd-176538&\cmark&\cite{feng2024crystal}\\
3cfea5f71d&A4B2C9\_aP30\_2\_4i\_2i\_9i:As-Gd-O&0.84&0.64&icsd-128943&\cmark&\cite{locke2023kenntnis}\\
f29f0ba1b5&A4B9C2\_aP30\_2\_4i\_9i\_2i:As-O-Pr&0.81&0.64&icsd-128942&\cmark&\cite{locke2023kenntnis}\\
694901187c&A2B21C6\_cF116\_225\_c\_afh\_e:La-Pd-Si&0.65&0.52&icsd-128830&\cmark&\cite{semeno2024ce2pd21si6}\\
b3b5b433f1&ABCD\_oP16\_62\_c\_c\_c\_c:H-La-Mg-Sn&0.59&0.93&icsd-168473&\cmark&\cite{yartys2024structure}\\
a8966b48b6&A12BC3D2\_hP36\_186\_4c\_b\_c\_ab:O-Rb-Sb-Se&0.59&0.61&icsd-87390&\cmark&\cite{robert2023syntheses}\\
8735f5f4ff&A5B12C2\_mC38\_12\_a2i\_2i2j\_h:Ho-O-Ru&0.51&0.82&icsd-176943&\cmark&\cite{patel2023hydrothermal}\\
d439a144f5&A9B3C2\_hP28\_194\_hk\_bf\_f:Br-Cs-Rh&0.84&0.47&icsd-175158&\xmark&\cite{liu2024exploring}\\
849128278d&A9B3C2\_hP28\_194\_hk\_bf\_f:Br-Rb-Rh&0.82&0.34&icsd-175162&\xmark&\cite{liu2024exploring}\\
6aa7e5dcc0&ABCD4\_oP28\_62\_a\_c\_c\_2cd:Ca-Ga-Gd-O&0.71&0.41&icsd-126832&\xmark&\cite{gai2024superior}\\
bdb468a870&A5B12C2\_mC38\_12\_a2i\_2i2j\_h:Dy-O-Ru&0.47&0.77&icsd-176939&\xmark&\cite{patel2023hydrothermal}\\
fd12110baa&A12B2C5\_mC38\_12\_2i2j\_h\_a2i:O-Ru-Tm&0.36&0.72&icsd-176942&\xmark&\cite{patel2023hydrothermal}\\
55e63b444a&A5B12C2\_mC38\_12\_a2i\_2i2j\_h:Lu-O-Ru&0.28&0.72&icsd-176944&\xmark&\cite{patel2023hydrothermal}\\
\hline\hline
\end{tabularx}
\end{table*}

To identify the promising subspaces, a Wyckoff encode-based existence probability evaluation model is proposed, as shown in Fig.~\ref{fig: workflow detail 1}(a). In this model, each Wyckoff encode is represented as a graph, where nodes correspond to Wyckoff orbitals. Through message passing and pooling, the Wyckoff graph is mapped into a latent representation, where a standard classification network predicts the existence probability of each latent representation.

A total of 49,601 synthesized structures and 105,117 theoretical structures collected from the MP database are used to train and evaluate the model. 
After removing duplicates and excluding Wyckoff encodes with $P1$ symmetry, the dataset consists of 48,724 encodes from synthesized structures, considered as positive samples representing subspaces where synthesizable structures exist, and 84,187 encodes from theoretical structures, treated as unlabeled samples.
Since explicit negative samples are unavailable, a positive-unlabeled (PU) learning approach~\cite{jang2020structure, bekker2020learning} is employed. During training, the positive and unlabeled samples are split into two sets: 80\% and 20\% for training and testing, respectively. In total, 56 classifiers are trained, and their outputs are averaged to obtain the final evaluation result. Further training details are provided in the SI Sec. II.

The model's performance on the test set is illustrated in Figs.~\ref{fig: workflow detail 1}(b-c).
For samples derived from synthesized structures, the model achieves an accuracy of 88\% when the existence probability exceeds 0.5.
In contrast, only 22\% of samples from theoretical structures surpass this threshold, highlighting the model’s effectiveness in pinpointing promising subspaces for synthesizable structures.

To further evaluate its generalization ability, 2,019 Wyckoff encodes derived from newly synthesized structures in the ICSD, which are excluded from model training, are used for independent validation. As illustrated in Fig.~\ref{fig: workflow detail 1}(d), the model achieves a high true positive rate accuracy of 84\%, highlighting its strong transferability.

\subsection{Structure-based screening}
Once the promising subspaces are determined, the structures generated by the derivation method within these subspaces are relaxed to local minima using a UMLP implemented in the Crystal Hamiltonian Graph Neural Network (CHGNet)~\cite{deng2023chgnet}.
Notably, the UMLP used in this work achieves a mean absolute error as low as 30 meV/atom. The lower-energy candidates are then screened, followed by a systematic similarity check to remove duplicates. The remaining unique structures are further relaxed using \textit{ab initio} methods, with density functional theory calculations performed using the Vienna \textit{ab initio} Simulation Package (VASP)~\cite{kresse1996efficient}. 
To assess thermodynamic stability, all relevant structures for each composition are retrieved from the MP database to evaluate the convex hull energy ($E_{hull}$). 

\begin{figure*}[ht]
    \centering
    \includegraphics[width=1\linewidth]{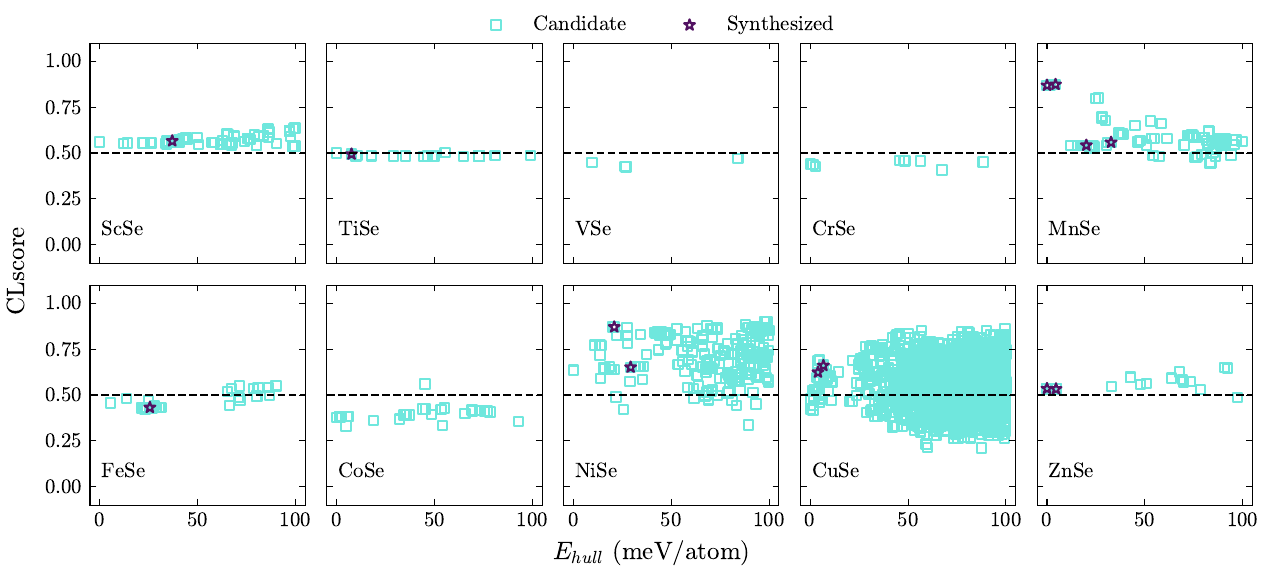}
    \caption{The energy and CLscore distribution of candidate structures for the XSe (X = Sc-Zn) compounds, where CLscore $>$ 0.5 indicates high synthesizability. Note that experimentally synthesized structures are marked by purple stars, while predicted structures are represented by cyan squares.}
    \label{fig: XSe}
\end{figure*}

\begin{figure*}[ht]
    \centering
    \includegraphics[width=1\linewidth]{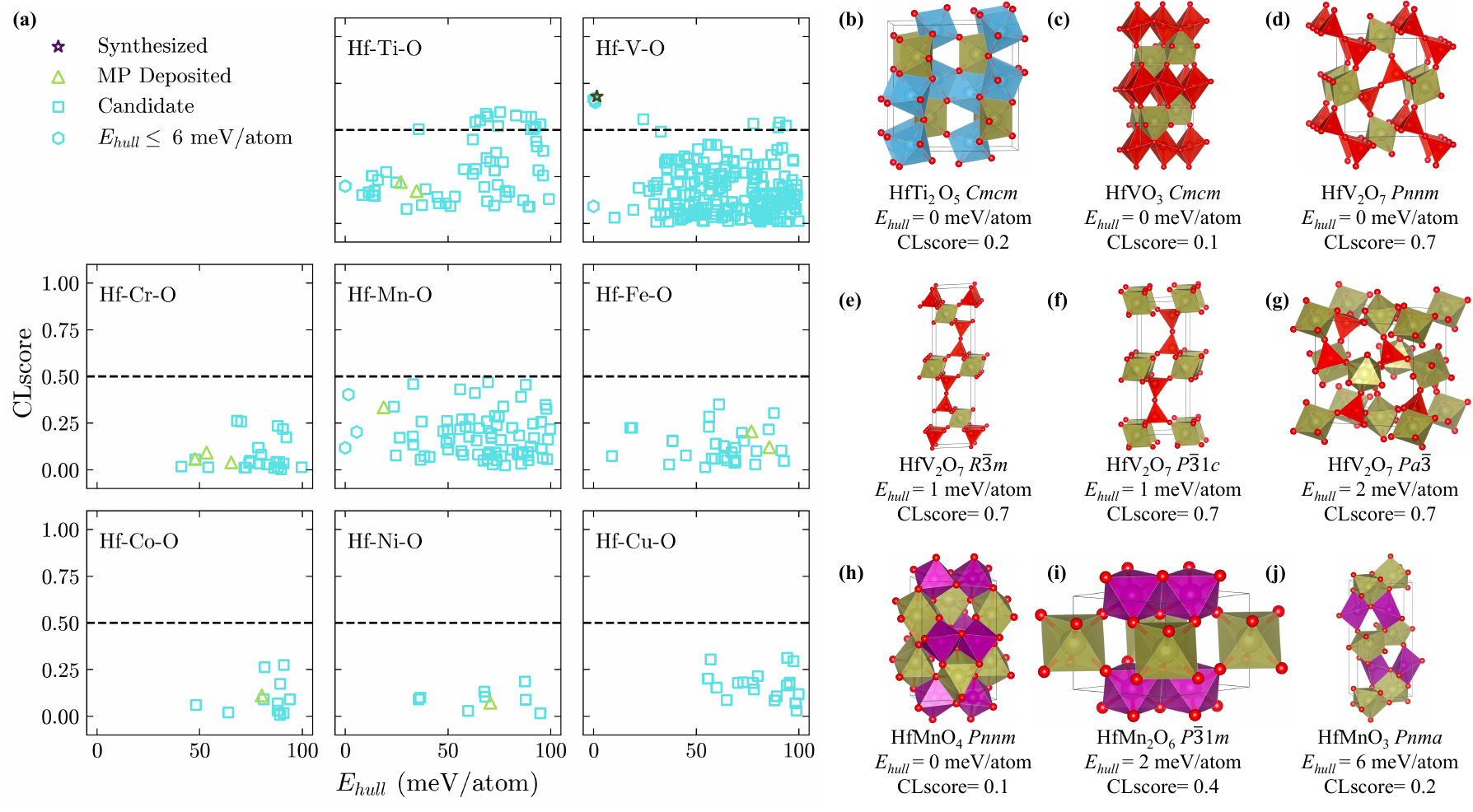}
    \caption{ (a) The energy and CLscore distribution of candidate structures for Hf-X-O (X = Ti-Cu) compounds. The synthesized structure of HfV$_2$O$7$ ($Pa\overline{3}$) is marked by a purple star, while green triangles represent theoretical structures from the MP database. Structures identified in this work are denoted by cyan squares and hexagons, with hexagons highlighting those with an $E_{hull}$ value of 6 meV/atom or lower. (b-j) Candidate structures showcasing distinct metal-oxygen coordination environments. Yellow polyhedra represent Hf-O local configurations, while blue, red, and purple polyhedra denote Ti-O (b), V-O (c-g), and Mn-O (h-j) coordination environments, respectively.
    }
    \label{fig: HfXO}
\end{figure*}

Candidates with \( E_{hull} \) below 100 meV/atom are further evaluated using our structure-based synthesizability model, fine-tuned on the PU-CGCNN with 4,052 newly synthesized structures collected from the ICSD between 2022 and 2024. The model outputs a crystal-likeness score (CLscore) to quantify the synthesizability of a given crystal structure, where CLscore $>$ 0.5 indicates that the structure is synthesizable.
The fine-tuned model achieves an accuracy of 91$\%$, slightly surpassing the performance of the previous model. The performance of the fine-tuned model on the stable structures from GNoME-based exploration~\cite{merchant2023scaling} is similar to that of the original model, with 63$\%$ of the stable structures predicted as synthesizable.
Details on the calculations and the fine-tuned model are provided in the SI Sec. III.

\section{Applications}
\subsection{Synthesizable structures in the GNoME dataset}

Recently, GNoME predicted that 554,054 novel structures lie within 1 meV/atom of the convex hull~\cite{merchant2023scaling}. To identify the potentially synthesizable structures from this dataset, we filter them using Wyckoff encode-based existence probability ($P > 0.5$) and structure-based synthesizability criteria (CLscore $> 0.5$). 
Specifically, when considering structural synthesizability alone, 347,833 structures are deemed synthesizable. However, only 92,310 of these structures satisfy both the conditions of $P > 0.5$ and CLscore $> 0.5$. The elemental distribution of these structures, in comparison to that of synthesized structures, is shown in Fig.~\ref{fig: gnome_icsd}(a). 
As illustrated, oxides are the most commonly synthesized compounds, while phosphides dominate the set of synthesizable structures in the GNoME dataset. Interestingly, the numbers of novel structures containing oxygen and phosphorus in the GNoME dataset are comparable, with 44,611 for oxides and 43,262 for phosphides. This trend aligns with previous studies, which have shown that synthesized phosphides exhibit greater thermodynamic stability than oxides ~\cite{sun2016thermodynamic}. Additionally, due to the chemical similarity of elements within the same group, the distribution of synthesizable structures reflects a similar trend, as seen in element sets like Sc–Y–La, Co–Rh–Ir, and Ni–Pd–Pt.


Further investigation reveals that when compared with the ICSD, seventeen structures in the GNoME dataset have been experimentally synthesized recently, as summarized in Table~\ref{tab: gnome_icsd}. To further assess the performance of our framework on these validated structures, we find that fourteen of them are located within Wyckoff-encoded subspaces associated with a high existence probability of synthesizable structures ($P > 0.5$), and fourteen exhibit CLscores exceeding 0.5. By combining the criteria from both the Wyckoff encode-based model and the fine-tuned PU-CGCNN, our framework successfully identifies 11 out of the 17 structures ($\sim$65\%) as synthesizable. These findings further demonstrate the effectiveness of our machine-learning models in predicting synthesizable structures.

We also analyzed the number of structures as a function of the number of constituent elements, comparing synthesized structures from the ICSD with synthesizable structures from the GNoME dataset, as shown in Fig.~\ref{fig: gnome_icsd}(b). Ternary and quaternary compounds are the most prevalent in both the synthesized structures from the ICSD and the synthesizable structures from the GNoME dataset. 
A notable distinction arises in quaternary compounds: only 23\% (3,076 out of 13,107) of synthesized structures contain three or more metallic elements, compared to 81\% (45,195 out of 55,619) among synthesizable quaternary structures in the GNoME dataset. Of these 45,195 structures, 39,067 feature at least one pair of metallic elements with atomic radius differences below 15\%, indicating the potential for forming substitutionally disordered compounds ~\cite{cheetham2024artificial}.
However, our current model is not capable of handling substitutionally disordered compounds. Given the limited availability of synthesized multi-nary ordered compounds and the challenges our models face in predicting disordered compounds ~\cite{simonov2020designing}, further development of our approach for disordered structures is crucial for advancing the understanding and design of complex multi-component material systems, such as high entropy alloys ~\cite{george2019high} and high entropy ceramics ~\cite{oses2020high}.

\subsection{XSe (X = Sc-Zn) compounds}

We select XSe (X = Sc-Zn) compounds from the MP dataset as ideal benchmarks to evaluate the model's predictive capability in distinguishing synthesizable materials from hypothetical candidates. Their intrinsic diversity, comprising 22 experimentally synthesized and 16 theoretically predicted structures, ensures a comprehensive assessment of the framework across different structure types.

Initially, possible candidate structures of XSe (X = Sc-Zn) are generated through structure derivation based on our structural prototype database, which contains 13,426 prototypes. To prevent combinatorial explosion, several constraints are imposed, including a maximum of 12 atoms per primitive cell, 6 orbitals, and a group-subgroup index limit of 10. This procedure yields a total of 102,858 unique structures per compound, which are systematically classified into 4,245 distinct subspaces, each characterized by a Wyckoff encode. 

Our Wyckoff encode-based evaluation model further identifies an average of 659 promising subspaces, effectively narrowing down the selection to approximately 25,035 derived structures per compound for further screening. After structural relaxation using UMLP and DFT, 323 structures per compound with \( E_{hull} < 100 \) meV/atom are selected for synthesizability evaluation. Detailed statistics for each compound are summarized in Table S3.
The distributions of synthesizability scores and formation energies obtained from DFT calculations for the screened structures are illustrated in Fig.~\ref{fig: XSe}. A comparison between the predicted structures and those in the MP database reveals that 13 synthesized structures with \( E_{hull} \) values below 100 meV/atom are successfully reproduced. Among these, 11 structures exhibit high synthesizability (CLscore $>$ 0.5), while the remaining two, with scores of 0.49 and 0.43, fall slightly below the synthesizability threshold, highlighting the effectiveness of our framework.

Note that our predictions reveal a pronounced scarcity of low-energy structures (\( E_{hull} < 50\) meV/atom)  with a synthesizability probability exceeding 0.5 for the VSe, CrSe, FeSe, and CoSe systems. 
Among XSe (X= Ti, V, Cr, Fe, Co) systems, nine failure cases, where our framework fails to reproduce the synthesized structures. A deeper analysis reveals that these failures primarily arise from the limited accuracy of the UMLP, which leads to the unintended removal of these candidates during the initial screening stage. While accurate DFT calculations could effectively rectify these inaccuracies, their prohibitively high computational cost severely limits their practicality for large-scale screening, the details are shown in Fig. S7 and Table. S4.

Beyond the experimentally synthesized structures, a series of low-energy structures with high synthesizability (CLscore $>$ 0.5) are identified in XSe (X = Sc, Ti, Mn, Ni, Cu, Zn), presenting promising candidates for experimental validation. 
These structures cluster around their synthesized counterparts, as shown in Fig.~\ref{fig: XSe}, indicating structural proximity within the configurational space.
For instance, two additional structures are identified near the synthesized ZnSe structures, all characterized by vertex-sharing Zn-Se tetrahedrons, as illustrated in Fig. S8. This structural resemblance suggests the potential feasibility of synthesizing these polymorphs through precise control of kinetic pathways.


\subsection{Hf-X-O (X=Ti-Cu) chemical systems}

Oxides incorporating heavy and magnetic elements exhibit unique electronic and magnetic properties, making them promising candidates for advanced functional materials~\cite{hoffmann2020magnetic}, as evidenced by the 
permanent magnetic BaFe$_{12}$O$_9$~\cite{de2021progress}, the multiferroic BiFeO$_3$~\cite{wang2003epitaxial,spaldin2019advances,fiebig2016evolution}, and superconducting La$_3$Ni$_2$O$_7$~\cite{ko2024signatures,sun2023signatures}.
However, previous studies have shown that oxides often exhibit metastability due to their exceptionally strong ionic bonds, which can stabilize energetically unfavorable atomic arrangements~\cite{sun2016thermodynamic, aykol2018thermodynamic}. Consequently, energy-driven CSP approaches pose significant challenges in identifying experimentally realizable materials within these complex systems. 

Here, we apply our synthesizability-driven CSP method to Hf-X-O (X = Ti-Cu) chemical systems, which serve as archetypal oxides incorporating heavy and magnetic elements. The structure derivation process is conducted under defined constraints, including a maximum of 24 atoms per primitive cell, a limit of six orbitals, and a subgroup index restricted to 10, while ensuring an oxygen stoichiometry of at least 0.5. According to our prototype database, a total of 2,477,961 structures and 132,667 unique Wyckoff encodes are generated for each chemical system. Evaluation using the Wyckoff encode-based model yields approximately 186,633 structures per chemical system. Further screening with UMLP and DFT refines the selection, identifying an average of 172 structures per chemical system with $E_{hull} < 100$ meV/atom, as summarized in Table S5. 

The results of the formation energy analysis and structural synthesizability evaluation are presented in Fig.~\ref{fig: HfXO}(a).  From an energetic perspective, 56 Hf-X-O (X = Ti-Ni) structures are predicted to be more energy-favorable than those previously reported in the MP database. In particular, 8 structures with $E_{hull}\leq$ 6 meV/atom are identified for the Hf-X-O (X = Ti, V, Mn) systems, as sketched in Fig.~\ref{fig: HfXO}(b-j). No stable structures with formation energies on the convex hull or CLscores exceeding 0.5 are identified in the Hf-X-O (X = Cr, Fe, Co, Ni, Cu) systems. These findings are consistent with the absence of recorded synthesized structures for these systems in the MP database, further validating the reliability of our framework. 

For HfV$_2$O$_7$, three newly identified structures, which are nearly degenerate in energy with the synthesized structure of $Pa\overline{3}$, exhibit high structural synthesizability scores of approximately 0.7.
In particular, these structures exhibit structural features closely resembling the experimental phase, characterized by vertex-sharing Hf-O octahedra and V-O tetrahedra. Each octahedron is connected to six tetrahedra, while each tetrahedron is linked to three octahedra and one additional tetrahedron. 
The detailed structural information is provided in the SI Sec. IV. 


Previous studies have demonstrated that the HfV$_2$O$_7$ phase with the $Pa\overline{3}$ structure can be synthesized via the reaction: HfO$_2$ + 2VO$_2$ + 1/2 O$_2$ $\rightarrow$ HfV$_2$O$_7$ 
with a reaction energy of approximately -91 meV/atom ~\cite{liu2021negative}. Given that the predicted structures are energetically quasi-degenerate with the experimental structure and share similar structural characteristics, they are likely accessible through synthesis using the same precursor, provided that the synthesis conditions are carefully tuned to navigate the kinetic pathway effectively. In addition, our DFT calculations identify an alternative synthesis route: HfO$_2$  + V$_2$O$_5$ $\rightarrow$ HfV$_2$O$_7$ with a reaction energy of -72~meV/atom, indicating that this pathway could serve as a viable approach for experimental realization.

It is noteworthy that temperature-induced phase transitions have been observed for the $Pa\overline{3}$ phase upon cooling from 400 K to 300 K, without significant structural changes~\cite{turquat2000structural,yamamura2011negative}. Given the high similarity in structural characteristics and energetic properties, the identified structures may be related to these experimentally observed phases. However, further experimental validation is required to assess their consistency with our predictions.


\section{Conclusion}
In this study, a synthesizability-driven CSP framework is developed to overcome the limitations of conventional energy-based approaches, which often fail to identify experimentally realizable metastable structures. Our methodology integrates symmetry-guided substitutional structure derivation, Wyckoff encode-based subspace filtering, and machine-learning-assisted structural screening to systematically identify low-energy candidates with high synthesizability. 
Benchmarking against XSe (X = Sc-Zn) compositions successfully reproduces
13 experimentally synthesized structures, validating the effectiveness of our approach. 
Applying this framework to the GNoME-identified candidates and Hf-X-O (X = Ti-Cu) systems leads to several promising candidates exhibiting favorable thermodynamic stability or high synthesizability. This work establishes a robust and efficient strategy for accelerating the discovery of synthesizable functional materials.

Despite the effectiveness of our framework in assessing the structural synthesizability of inorganic materials, the synthesis process remains inherently complex and presents significant challenges, as it depends on precursor selection, reaction kinetics, and the optimization of synthesis conditions such as temperature and pressure. Future research should focus on developing predictive models that integrate these factors to enable more comprehensive synthesis planning, enhancing the practical applicability of crystal structure prediction, and bridging the gap between theoretical predictions and experimental realization.

\section*{Acknowledgments}
This research is supported by the National Natural Science Foundation of China under Grants Nos. T2225013, 12034009, 12174142, and 42272041, the National Key R\&D Program of China (Grant No. 2022YFA1402304), the Program for JLU Science and Technology Innovative Research Team, Program for Jilin University Computational Interdisciplinary Innovative Platform. Part of the calculation is performed in the high-performance computing center of Jilin University.

\bibliography{bibs}
\end{document}